\documentclass[12pt]{article}

\usepackage[a4paper,left=30mm,right=20mm,top=20mm,bottom=20mm]{geometry}
\usepackage{graphics}
\usepackage{amsmath}
\usepackage{amssymb}
\usepackage{epsfig}
\usepackage{cite}
\usepackage{xcolor}
\usepackage[unicode]{hyperref}
\newcommand{\bea}{\begin{eqnarray}}
\newcommand{\eea}{\end{eqnarray}}
\newcommand{\be}{\begin{equation}}
\newcommand{\ee}{\end{equation}}

\title{Forward $J/\psi+J/\psi$ and $J/\psi+\psi'$ production with High Energy Factorization}

\author{S.P.~Baranov$^1$, A.V.~Lipatov$^{2,3}$, M.A.~Malyshev$^{2,4}$, \\ A.A.~Prokhorov$^{3}$, P.M.~Zhang$^5$}

\begin{document}

\maketitle

\begin{center}

{\it $^{1}$P.N.~Lebedev Institute of Physics, Moscow 119991, Russia}\\
{\it $^{2}$Skobeltsyn Institute of Nuclear Physics, Lomonosov Moscow State University, Moscow 119991, Russia}\\
{\it $^{3}$Joint Institute for Nuclear Research, Dubna 141980, Moscow region, Russia}\\
{\it $^{4}$Moscow Aviation Institute, Moscow 125993, Russia}\\
{\it $^{5}$School of Physics and Astronomy, Sun Yat-sen University, Zhuhai 519082, China}

\end{center}

\vspace{0.5cm}

\begin{center}

{\bf Abstract }
       
\end{center}

\indent

We calculate the cross sections of associated $J/\psi + \psi^\prime$ 
and $J/\psi + J/\psi$
production in $pp$ 
collisions at $\sqrt s = 13$~TeV in the forward kinematic region.
The High Energy Factorization ($k_T$-factorization) framework
supplemented with the Catani-Ciafaloni-Fiorani-Marchesini evolution of gluon densities in a proton is applied.
We demonstrate that latest data on $J/\psi + J/\psi$
production and 
first experimental data on $J/\psi + \psi^\prime$ events taken very recently by the LHCb Collaboration
can be described well by the color singlet terms and contributions 
from the double parton scattering (DPS) with the standard
choice for $\sigma_{\rm eff}$ parameter.
The relative production rate $\sigma(J/\psi + \psi^\prime)/\sigma(J/\psi + J/\psi)$
is found to be sensitive to the DPS terms
as well as to feeddown contributions.

\vspace{1.0cm}


\newpage 

\indent

Inclusive and associated production of quarkonium states in hadron-hadron
collisions at high energies attracts much attention from both theoretical 
and experimental sides.
It probes Quantum Chromodynamics (QCD) in perturbative and non-perturbative 
regimes and provides important information about the interaction dynamics.
New data, as soon as they appear, trig theoretical activity aimed at their
understanding and description. Our present note is motivated by the first
measurement\cite{JPsiPsi-LHCb-13} of the associated $J/\psi+\psi^\prime$ 
production 
and very recent data\cite{JPsiJPsi-LHCb-13} on $J/\psi+J/\psi$
production
in $pp$ collisions taken by the LHCb Collaboration in the 
forward kinematic region at $\sqrt s = 13$~TeV. These data include
contributions from both 
single and double parton scattering production mechanisms (SPS and DPS, respectively).
The data on $J/\psi+\psi^\prime$ events 
present the differential cross sections as functions of several kinematic 
variables, namely, the rapidity and azimuthal angle differences between the
$J/\psi$ and $\psi^\prime$ mesons 
$\Delta y(J/\psi, \psi^\prime)$ and $\Delta \phi(J/\psi, \psi^\prime)$, the
transverse momentum $p_T(J/\psi, \psi^\prime)$, rapidity $y(J/\psi, \psi^\prime)$
and invariant mass $M(J/\psi, \psi^\prime)$ of the $J/\psi+\psi^\prime$ system.
Similar observables as well as transverse momentum, 
rapidity of either $J/\psi$ meson and
transverse momentum asymmetry ${\cal A}_T$ of the two $J/\psi$ mesons
have been investigated in the latest
$J/\psi+J/\psi$ analysis\cite{JPsiJPsi-LHCb-13}. 
Moreover, the relative production rate $\sigma(J/\psi + \psi^\prime)/\sigma(J/\psi + J/\psi)$
has been reported\cite{JPsiPsi-LHCb-13} for the first time.

A commonly accepted framework for the description of heavy quarkonia 
production and decays is provided by non-relativistic QCD (NRQCD) 
approximation\cite{NRQCD-1,NRQCD-2}.
Explicit calculations\cite{DoubleJPsi-Lansberg2} show that the main role in 
the production of quarkonium pairs at forward rapidities is played by the color 
singlet (CS) mechanism, whereas the color octet (CO) contributions are much smaller.
The tree-level next-to-leading order (NLO$^*$) CS calculations for SPS mechanism
performed with the \textsc{helac-onia} tool\cite{DoubleJPsi-Gen1, DoubleJPsi-Gen2} 
tend to overestimate the LHCb data\cite{JPsiPsi-LHCb-13}, especially at low 
$\Delta y(J/\psi, \psi^\prime)$, low $M(J/\psi, \psi^\prime)$ and moderate 
$p_T(J/\psi, \psi^\prime)$, though still remain consistent with them within the 
large theoretical uncertainties.
The latest data\cite{JPsiJPsi-LHCb-13} on forward $J/\psi+J/\psi$
events can also be described by the NLO$^*$ CS calculations
within the large uncertainties, although the DPS
contributions were not subtracted from the measurements\footnote{The leading-order (LO) NRQCD predictions $J/\psi + J/\psi$ production 
are known\cite{DoubleJPsi-Kniehl1} and the NLO$^*$ contributions to the CS and CO 
mechanisms have been calculated\cite{DoubleJPsi-Lansberg1}.}.

In the present note we analyse the LHCb data\cite{JPsiPsi-LHCb-13, JPsiJPsi-LHCb-13} on 
double charmonia production
in the framework of the $k_T$-factorization\cite{kt-factorization} or, 
equivalently, the High Energy Factorization\cite{HighEnergyFactorization} formalism.
This formalism has certain technical advantages in the ease of including 
higher-order radiative corrections that can be taken into account in the form of
transverse momentum dependent (TMD) gluon distributions in a proton\footnote{See, 
for example, review\cite{TMD-Review} for more information.}.
Recently, it has been successfully applied to the double $J/\psi$ production
at the LHC\cite{DoubleJPsi-we1, DoubleJPsi-we2, DoubleJPsi-we3}. In particular, 
it was demonstrated\cite{DoubleJPsi-we2, DoubleJPsi-we3} that the early
data on $J/\psi + J/\psi$ events collected at $\sqrt s = 7$ and $13$~TeV at forward 
rapidities can be described well by the sum of color-singlet SPS and DPS mechanisms. 
It was also confirmed that the CO contributions at the LHC conditions can be safely 
omitted in the region of small invariant masses $M(J/\psi,J/\psi)$ in the forward rapidity region.
Now, in addition to $J/\psi$ pairs,
we consider 
$J/\psi+\psi^\prime$ production.
Our 
study is even more stimulated by the fact that previous 
theoretical attempts\cite{DoubleJPsi-Lansberg1, DoubleJPsi-Gen1, DoubleJPsi-Gen2, DoubleJPsi-Kniehl2} 
were not fairly successful.

\begin{figure}
\begin{center}
{\includegraphics[width=1.0\textwidth]{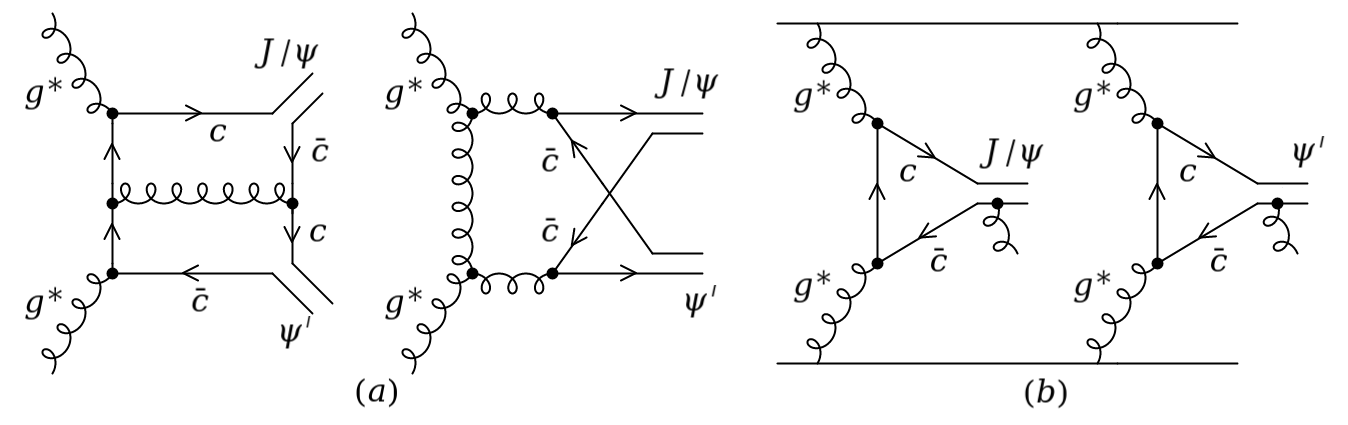}}\hfill
\caption{Examples of Feynman diagrams, contributing to the CS production of charmonium pairs.}
\label{fig:1}
 \end{center}
\end{figure}



We preserve full consistency with our previous 
investigations\cite{DoubleJPsi-we1, DoubleJPsi-we2, DoubleJPsi-we3} 
and employ the $k_T$-factorization QCD 
approach\cite{kt-factorization, HighEnergyFactorization}, as it was mentioned above.
This approach is based on the Balitsky-Fadin-Kuraev-Lipatov (BFKL)\cite{BFKL} 
or Catani-Ciafaloni-Fiorani-Marchesini (CCFM)\cite{CCFM} gluon evolution 
equations and can be used as a convenient alternative to explicit higher-order pQCD calculations. 

In the NRQCD framework, the colliding gluons produce heavy quark-antiquark pairs which further evolve into real
mesons. At forward rapidities, the leading CS contribution 
to associated $J/\psi + J/\psi$ or $J/\psi + \psi^\prime$ production
is represented by an
$\mathcal{O}(\alpha^4_s)$ partonic subprocess
\begin{gather}
  g^*(k_1) + g^*(k_2) \to c\bar c \left[^3S_1^{(1)}\right](p_1) + c\bar c \left[^3S_1^{(1)}\right](p_2),
  \label{CS-1}
\end{gather}
\noindent
where the four-momenta of all particles are indicated in the parentheses. 
This subprocess 
includes also feeddown contributions (namely, $J/\psi +\psi^{\prime}$ and/or 
$\psi^{\prime}+\psi^{\prime}$ production), where  
excited charmonium state $\psi^{\prime}$ decays into $J/\psi$ meson.
Other subleading subprocesses which involve the production of intermediate $P$-wave states and/or additional gluons (see\cite{DoubleJPsi-we3} for more details) were found 
to be small and not taken into account here.
In accordance with the $k_T$-factorization, the
initial gluons have nonzero transverse momenta $k_1^2 \equiv - {\mathbf k}_{1T}^2 \neq 0$, $k_2^2 \equiv - {\mathbf k}_{2T}^2 \neq 0$ and are off-shell.
Some examples of the relevant Feynman diagrams are shown in Fig.~\ref{fig:1}.


The gauge invariant expression for off-shell (depending on the 
virtualities of incoming gluons) production amplitude~(\ref{CS-1})
has been calculated earlier\cite{DoubleJPsi-we1}.
It contains spin and color projection 
operators\cite{ProjectionOperators-1, ProjectionOperators-2, ProjectionOperators-3, ProjectionOperators-4, ProjectionOperators-5} 
which guarantee the proper quantum numbers 
of the final state charmonia. In the NRQCD approximation which we are using 
the meson mass must be strictly equal to the sum of 
the constituent quark masses. However, this cannot be fulfilled simultaneously 
for both $J/\psi$ and $\psi^\prime$. So that, for the associated 
$J/\psi + \psi^\prime$ 
production we used a compromise value
$m_c= (m_{J/\psi}+m_{\psi'})/4$ (see discussion below). In all other respects our present theoretical
scheme is identical to that used in previous studies\cite{DoubleJPsi-we1, DoubleJPsi-we2, DoubleJPsi-we3}.
We only note that the initial gluon spin density matrix 
is taken in the form of so called "non-sense gauge", namely,
$\sum \epsilon^{\mu}\epsilon^{*\nu} = \mathbf{k}^{\mu}_T\mathbf{k}^{\nu}_T/\mathbf{k}_T^2$.
This expression converges to the ordinary 
$ -g^{\mu\nu}/2$ in the collinear limit $\mathbf{k}_T \to 0$ after averaging over the azimuthal angle. 

According to the $k_T$-factorization 
prescription\cite{kt-factorization, HighEnergyFactorization}, the 
contribution from CS production mechanism~(\ref{CS-1}) to $J/\psi + \psi^\prime$ cross section
is calculated as a convolution of the 
corresponding off-mass shell production amplitudes and TMD gluon densities in a proton:
\begin{gather}
  \sigma(pp \to J/\psi + \psi^\prime + X) = \int {1\over 16 \pi (x_1 x_2 s)^2} |{\cal \bar A}(g^* + g^* \to J/\psi + \psi^\prime)|^2 \times \nonumber \\
  \times f_g(x_1,\mathbf k_{1T}^2,\mu^2) f_g(x_2,\mathbf k_{2T}^2,\mu^2) d{\mathbf k}_{1T}^2 d{\mathbf k}_{2T}^2 d{\mathbf p}_{1T}^2 dy_1 dy_2 {d\phi_1\over 2\pi} {d\phi_2\over 2\pi} {d\psi_1\over 2\pi},
  \label{xsection}
\end{gather}
\noindent 
where $\psi_1$ is the azimuthal angle of the outgoing $J/\psi$ meson,
$\phi_1$ and $\phi_2$ are the azimuthal angles of the initial off-shell gluons having the 
longitudinal momentum fractions $x_1$ and $x_2$, 
$y_1$ and $y_2$ are the center of mass rapidities of the produced particles.
Note that the cross section of associated $J/\psi + J/\psi$ production can be calculated
in a similar way.
Here $f_g(x,{\mathbf k}_{T}^2, \mu^2)$ is the TMD 
gluon density in a proton taken at the scale $\mu^2$.
For the latter, we have tried two recent sets, referred to as JH'2013 set 2\cite{JH2013} and LLM'2022\cite{LLM-2022}. 
These gluon densities have been  obtained from a numerical solution of the CCFM equation and are now widely used in phenomenological applications
(see, for example,\cite{Motyka-photon, LMJ-PP, LM-Higgs, LLM-FL, LLM-photon} and references therein). 
The parameters of (rather empirical) input distributions employed in the JH'2013 gluon 
were derived from a fit to the HERA data on the proton structure functions $F_2(x,Q^2)$ 
and $F_2^c(x, Q^2)$ at small $x$. An analytical expression for the input gluon density 
in the very recent LLM'2022 set was suited to the best description of the LHC data 
on the charged hadron production at low transverse momenta 
in the framework of modified soft quark-gluon string model\cite{ModifiedSoftQuarkGluonStringModel-1, ModifiedSoftQuarkGluonStringModel-2} 
with taking into account the gluon saturation effects important at low scales. 
The necessary phenomenological parameters were deduced from the LHC and HERA data 
on several hard QCD processes (see\cite{LLM-2022} for more information). Both the JH'2013 set 2 and LLM'2022 
gluon distributions are available from the popular \textsc{tmdlib} package\cite{TMDLib2} and Monte-Carlo generator \textsc{pegasus}\cite{PEGASUS}.

\begin{figure}
\begin{center}
{\includegraphics[width=.32\textwidth]{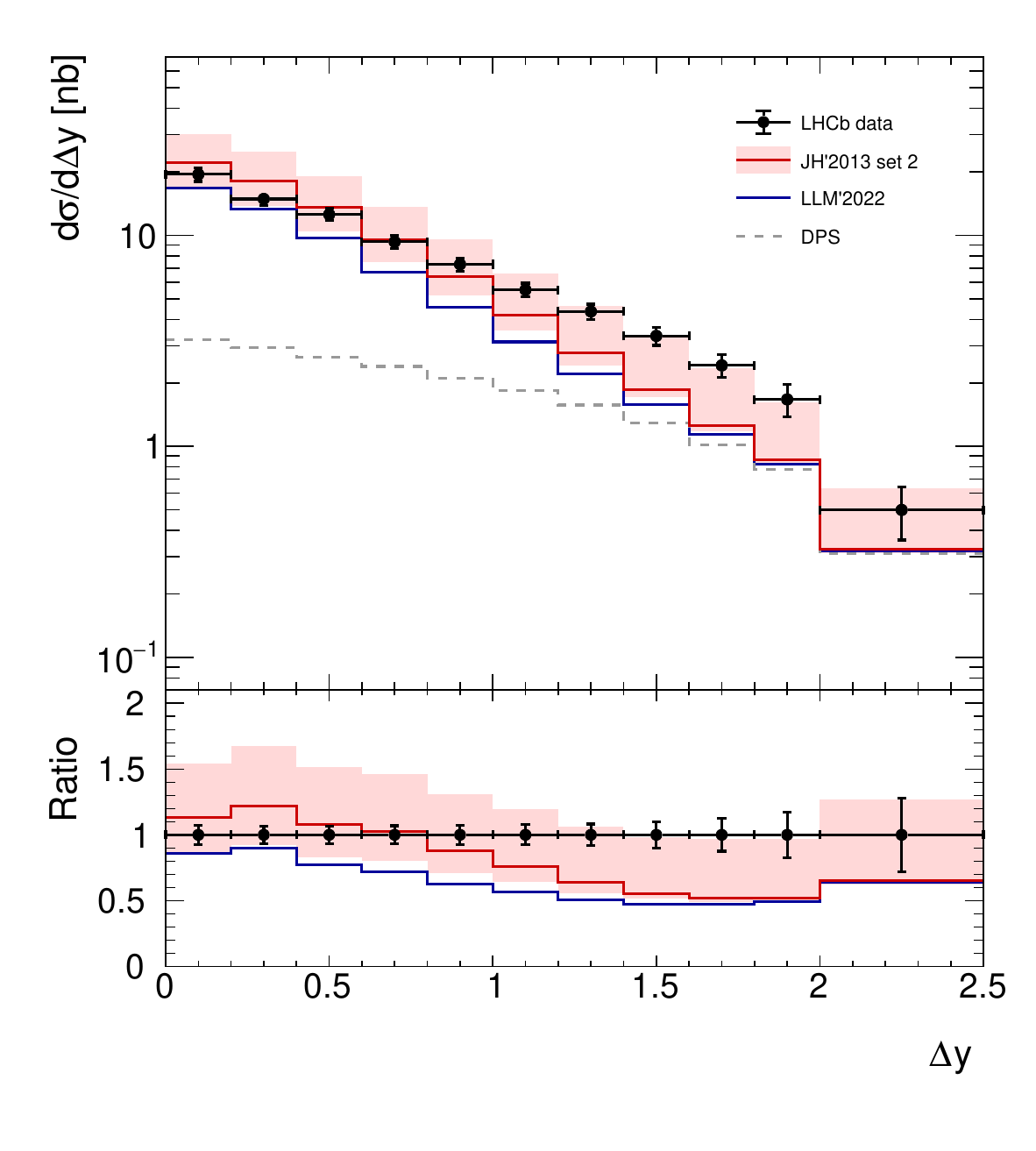}}
{\includegraphics[width=.32\textwidth]{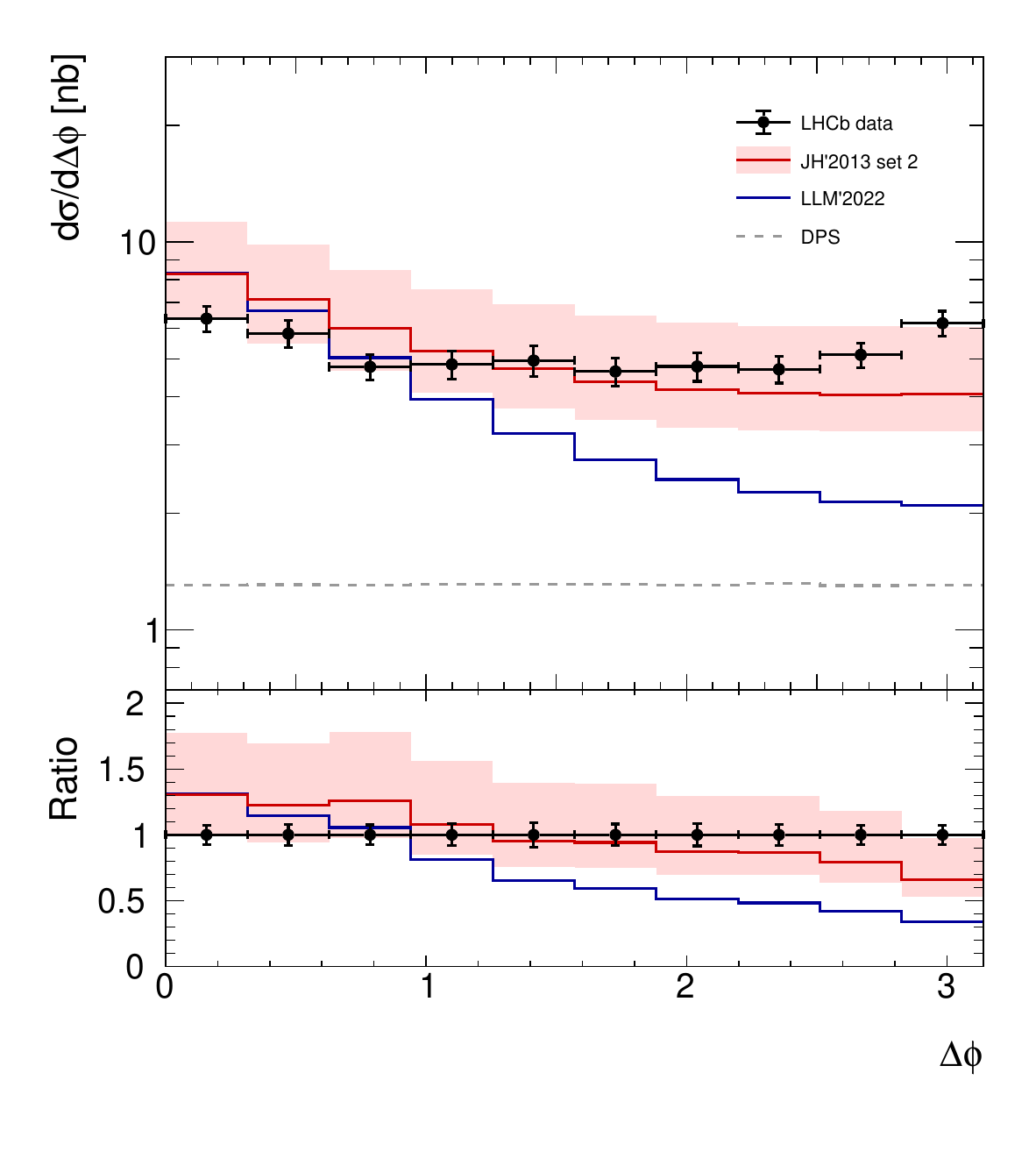}}
{\includegraphics[width=.32\textwidth]{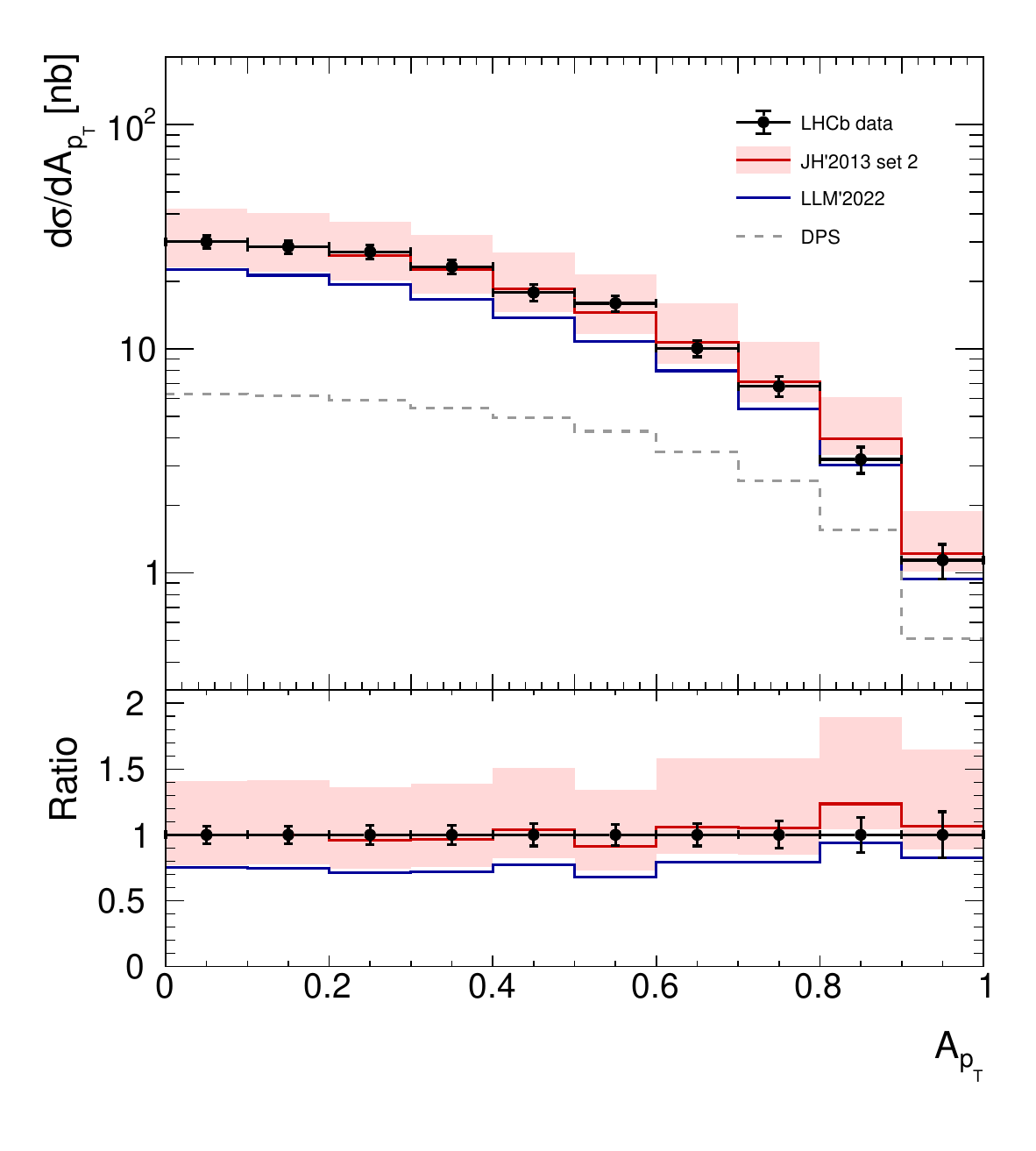}}
{\includegraphics[width=.32\textwidth]{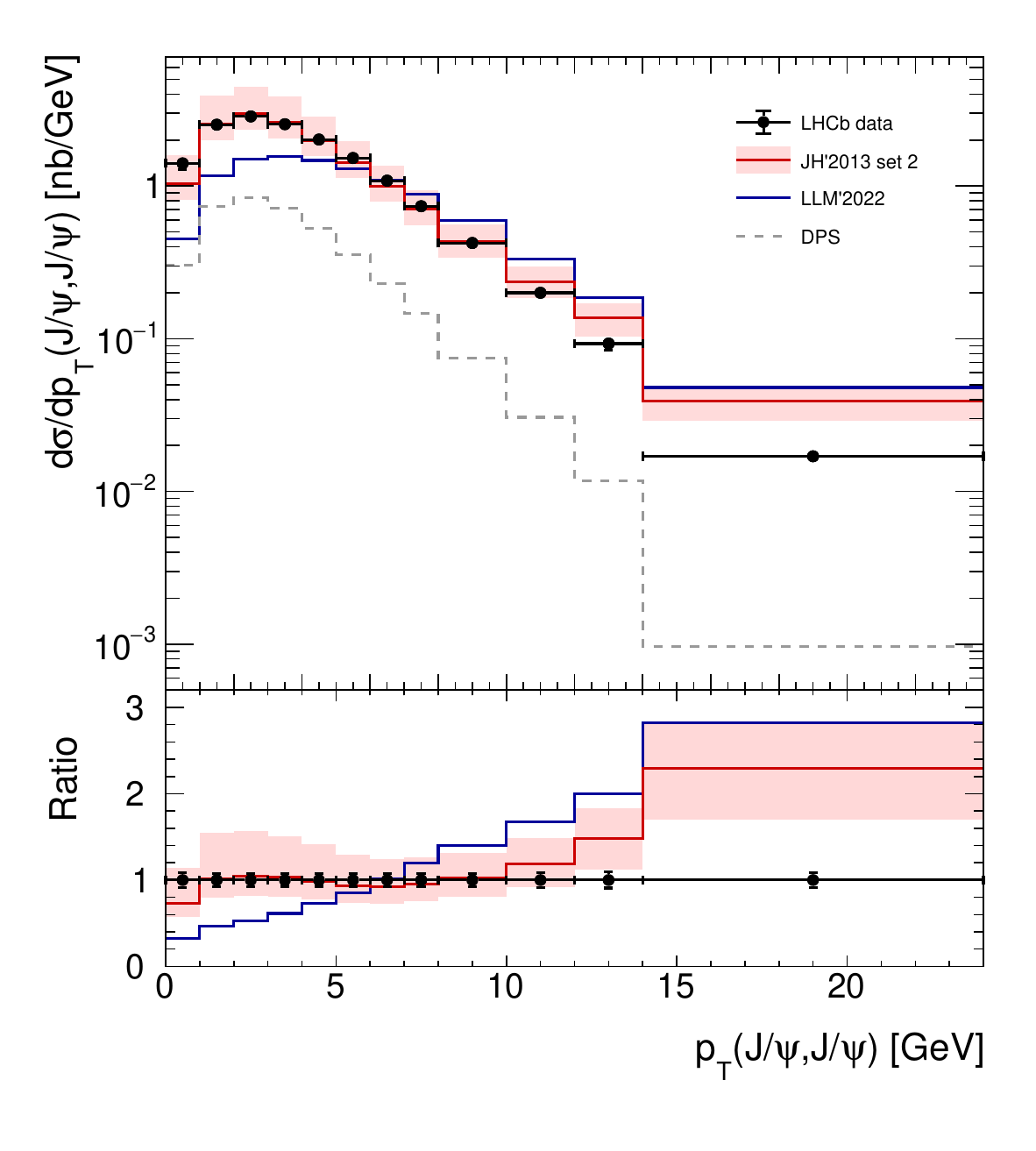}}
{\includegraphics[width=.32\textwidth]{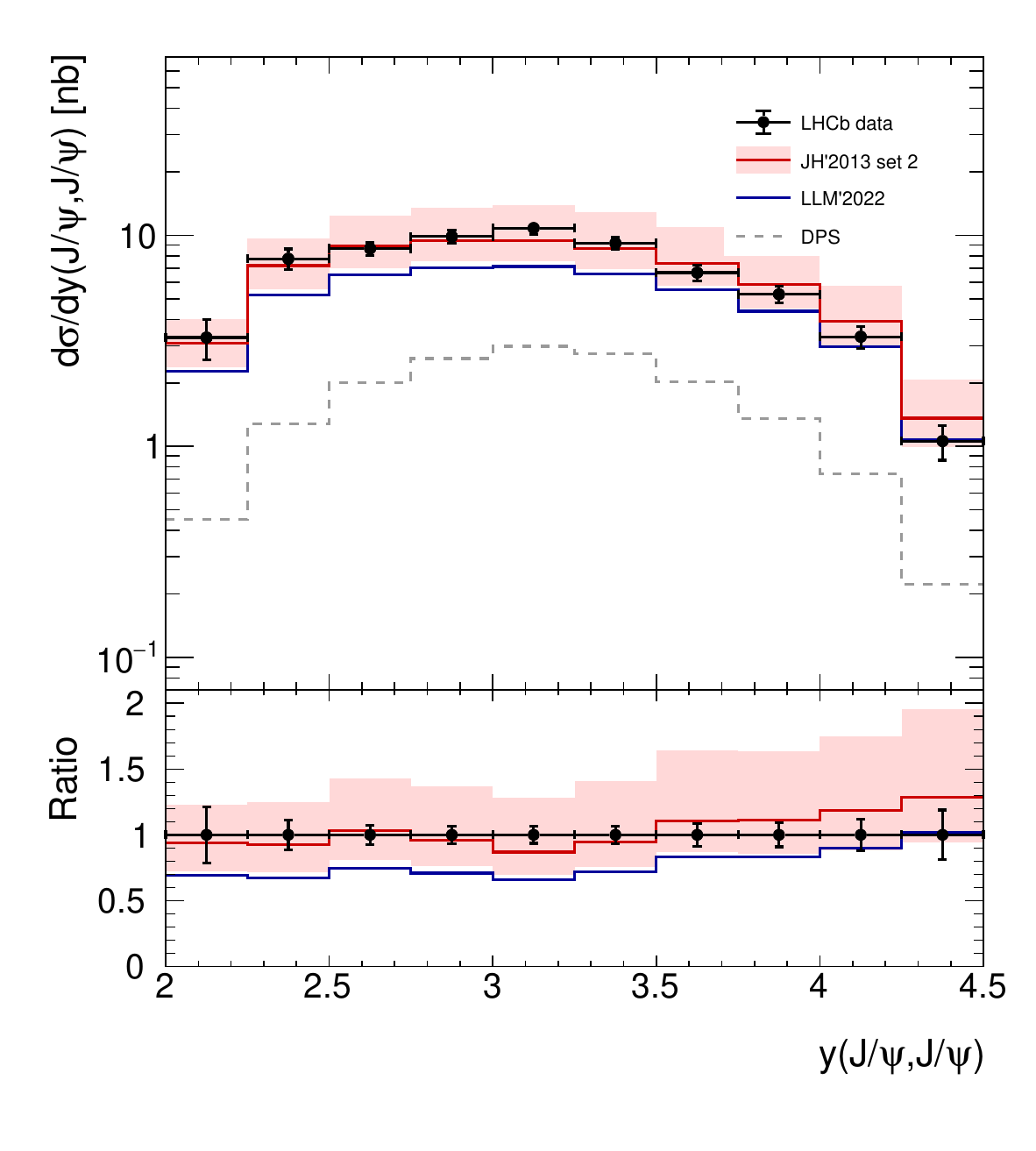}}
{\includegraphics[width=.32\textwidth]{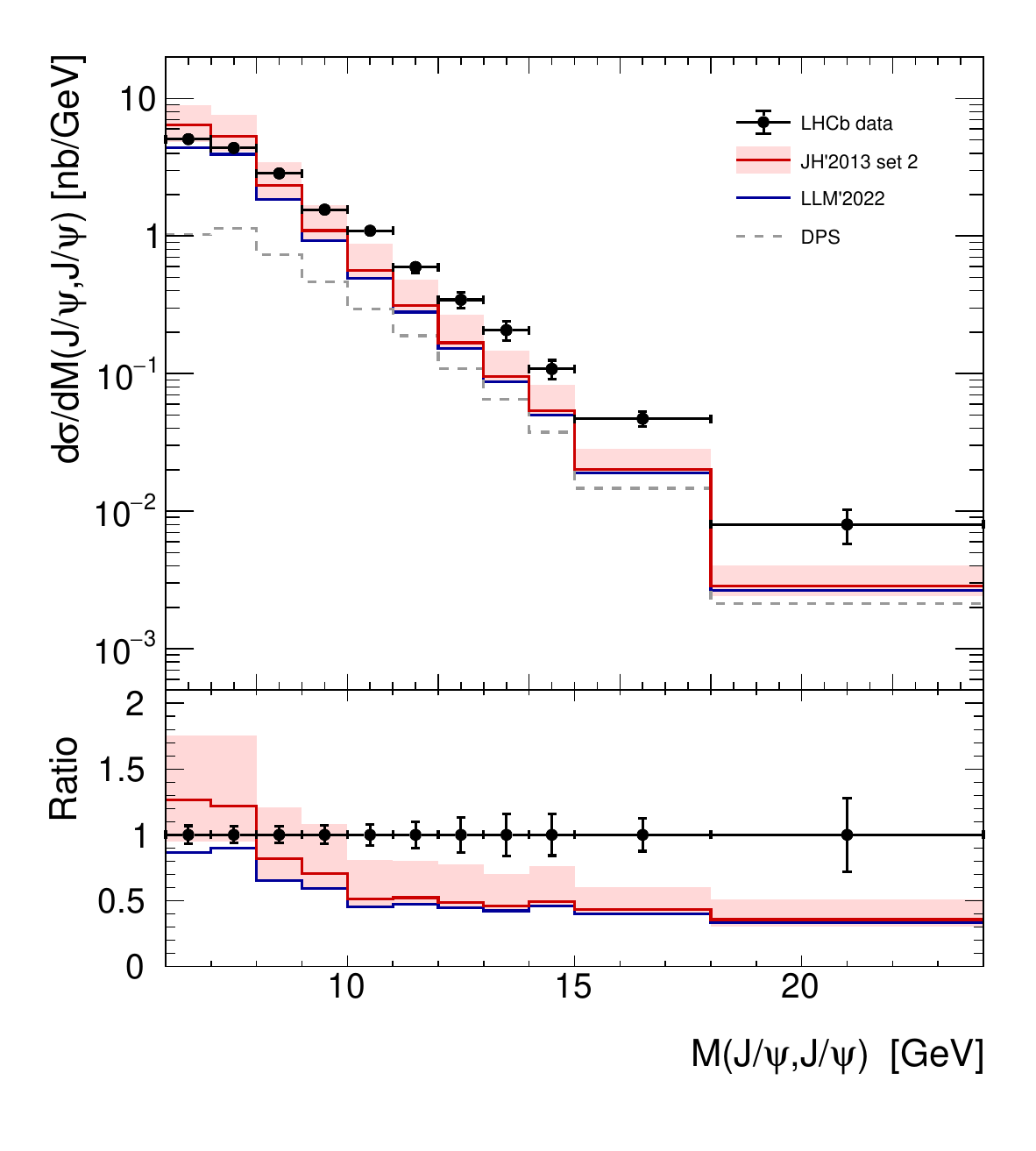}}
{\includegraphics[width=.32\textwidth]{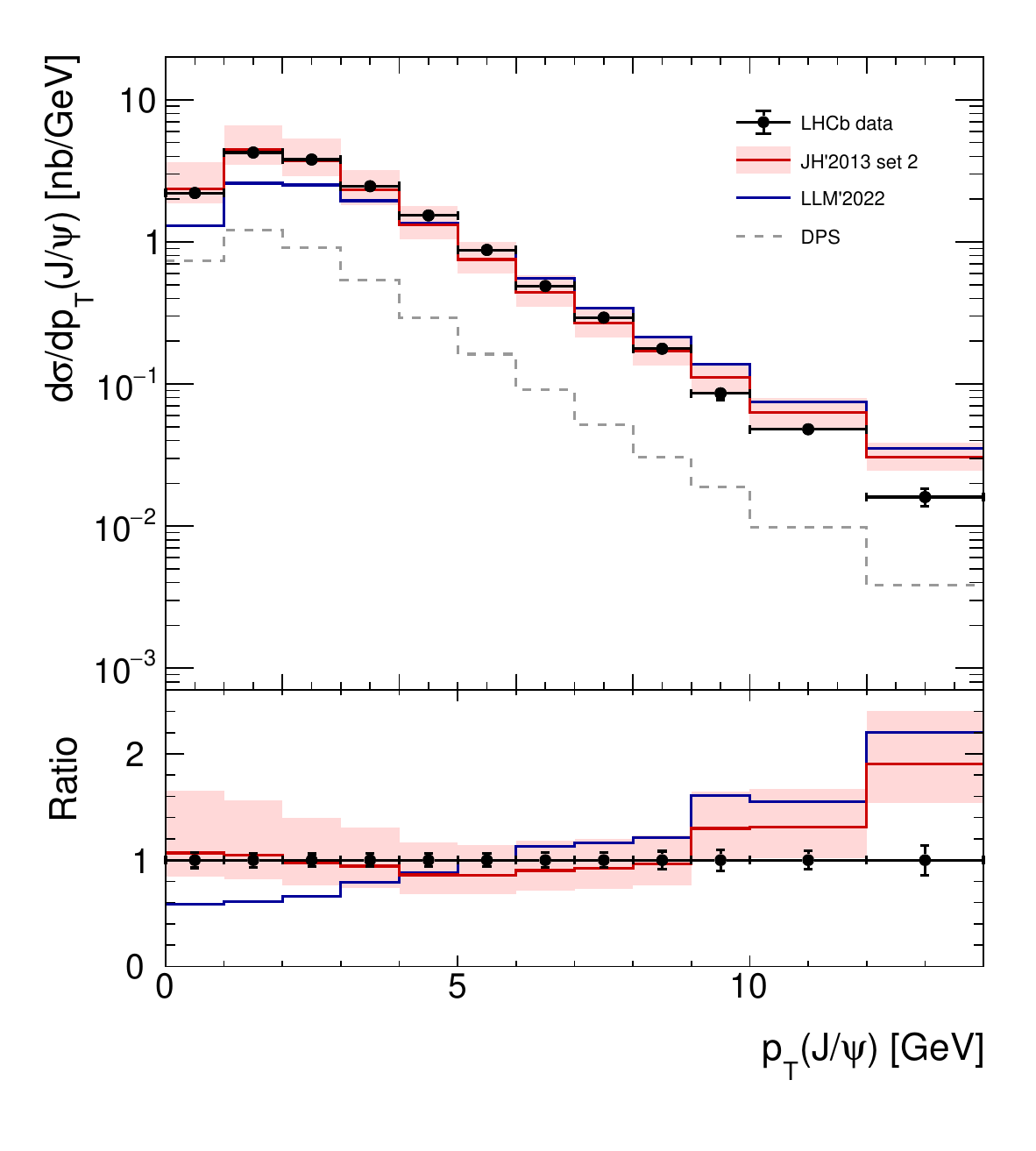}}
{\includegraphics[width=.32\textwidth]{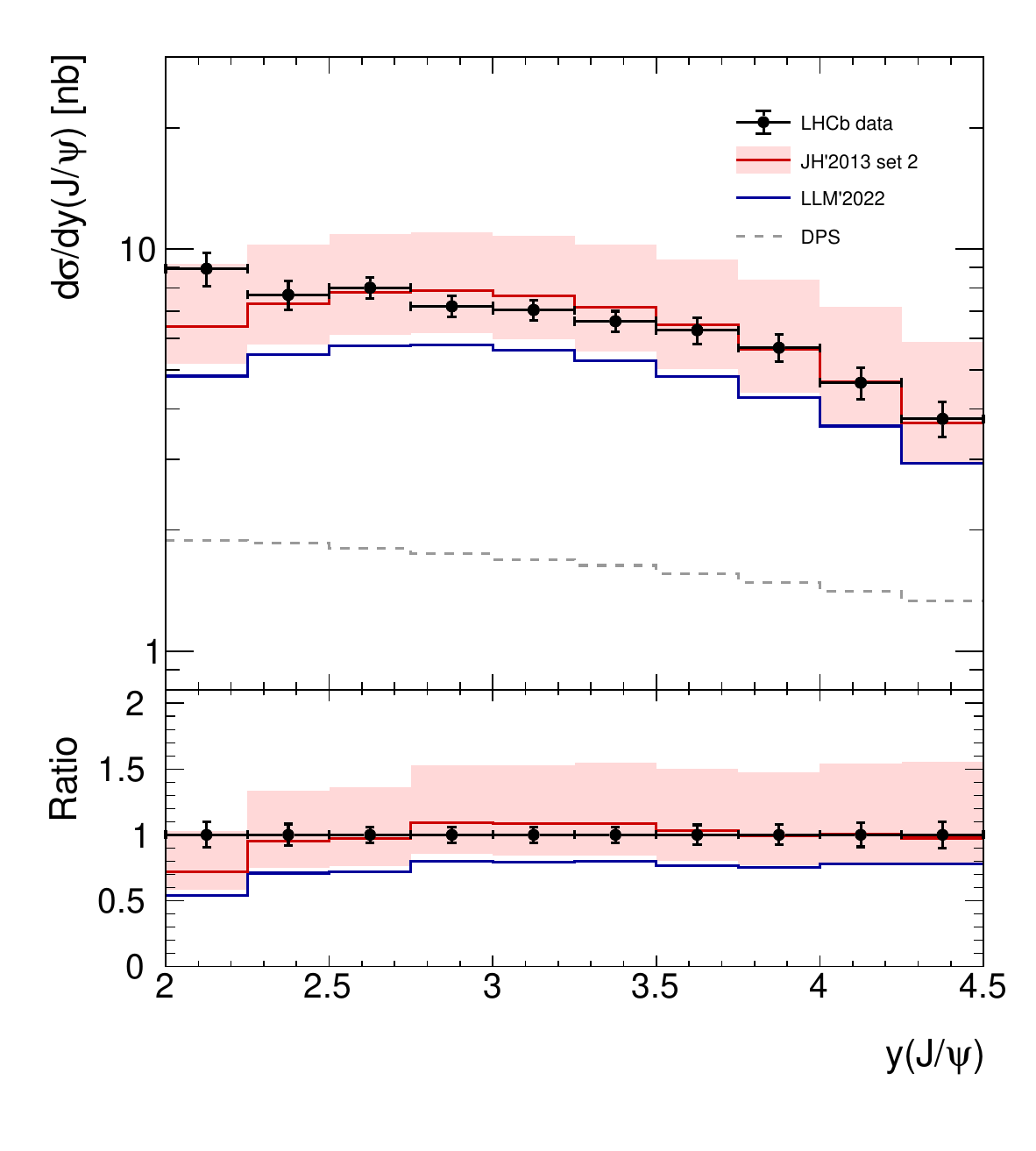}}
\caption{Differential cross sections of forward $J/\psi + J/\psi$ production at $\sqrt s = 13$~TeV as functions of their rapidity 
difference $\Delta y(J/\psi, J/\psi)$, azimuthal angle difference $\Delta\phi(J/\psi, J/\psi)$, 
transverse momentum assymetry ${\cal A}_T$, transverse momentum $p_T(J/\psi, J/\psi)$, 
rapidity $y(J/\psi, J/\psi)$, invariant mass $M(J/\psi, J/\psi)$ and 
transverse momentum $p_T(J/\psi)$ and rapidity $y(J/\psi)$ of either of $J/\psi$ meson.
Separately shown contrbutions from the DPS production mechanism.
The data are of LHCb\cite{JPsiJPsi-LHCb-13}.}
\label{fig:ATLAS_jpsi} 
 \end{center}
\end{figure}

In addition to the above, we also take into account the contributions from DPS
mechanism, which is now widely discussed in the literature.
In contrast with SPS subprocess~(\ref{CS-1}), where every quarkonium pair is produced 
in a single gluon-gluon collision, the DPS events originate from two independent parton interactions and are expected to be important at forward rapidities 
(see, for example,\cite{DPS-forward1, DPS-forward2, DPS-forward3}).
We apply a commonly used factorization formula:
\begin{gather}
  \sigma_{\rm DPS}(pp \to {\cal Q}_1 + {\cal Q}_2 + X) = {m \over 2} {\sigma(pp \to {\cal Q}_1 + X) \sigma(pp \to {\cal Q}_2 + X) \over \sigma_{\rm eff}},
  \label{DPS}
\end{gather}
\noindent 
where ${\cal Q}_{i} = J/\psi$, $\psi^\prime$ or $\chi_{cJ}$ and
$\sigma_{\rm eff}$ is the effective cross 
section which represents the degree of overlap in the transverse space
between two partonic interactions that constitute the DPS process. 
Note that $m = 1$ if ${\cal Q}_1 = {\cal Q}_2$ and $m = 2$ if ${\cal Q}_1$ and ${\cal Q}_2$ are different,
thus preventing double counting for identical particles.
The effective DPS cross section can be considered as a normalization constant which incorporates all "DPS unknowns"
into a single phenomenological parameter.
The derivation of formula~(\ref{DPS}) relies on two approximations: 
first, the double parton distribution function can be decomposed into longitudinal and transverse components and, second, the
longitudinal component reduces to the diagonal product of two independent single parton
densities. The latter is generally acceptable for the conditions of LHCb experiment.
The inclusive cross sections 
$\sigma(pp \to J/\psi + X)$ and/or $\sigma(pp \to \psi^\prime + X)$ 
involved in~(\ref{DPS}) are calculated within the NRQCD framework
supplemented with $k_T$-factorization (see, for example,\cite{TMD-charmonia} and references therein).
The effective cross section is chosen as $\sigma_{\rm eff} = 15$~mb, which is a
commonly accepted value\footnote{A very close value $\sigma_{\rm eff} = 13.8$~mb 
has been extracted\cite{DoubleJPsi-we2} from a fit to the LHCb data on the double $J/\psi$ production.
However, some other works on the double quarkonia production may seem to favor much 
smaller values: $\sigma_{\rm eff}\sim 5$~mb\cite{DPS-DQ1,DPS-DQ2,DPS-DQ3,DPS-DQ4}. 
We argue that this could be due to an incomplete treatment of the SPS contributions, see\cite{DoubleJPsi-we2} for more details.}.
Our calculation incorporates all 
possible feeddown contributions to the DPS cross section 
except the $\chi_{c0}$ production due to its low branching fraction to the $J/\psi$.

The meson masses were taken in our calculations 
as $m(J/\psi) = 3.096$~GeV, $m(\chi_{c1}) = 3.511$~GeV, $m(\chi_{c2}) = 3.556$~GeV and 
$m(\psi^{\prime}) = 3.686$~GeV (see\cite{PDG}). A complete list of used CS and CO LDMEs can be found\cite{DoubleJPsi-we3}. 
The branching fractions are ${Br}(\chi_{c1} \to J/\psi + \gamma) = 0.339$, ${Br}(\chi_{c2} \to J/\psi + \gamma) = 0.192$ 
and ${Br}(\psi^{\prime} \to J/\psi + X) = 0.529$.
The renormalization 
and factorization scales were calculated as $\mu_R^2 = \mu_F^2=\xi^2(\hat s+\mathbf Q_T^2)$, where $\mathbf Q_T$ is the total 
transverse momentum of the initial off-shell gluon pair. 
This choice observes full consistency with our previous work\cite{DoubleJPsi-we2}
and is dictated by the CCFM evolution. It comes from the kinematics of
gluon radiation, namely from the gluon angular ordering condition. 
The quantity $\hat s + \mathbf Q_T^2$ describes the phase space available for
the (angular ordered) gluon radiation and sets the upper bound both for the emitted 
and exchanged transverse momentum (see\cite{CCFM} for more information).
We use a two-loop formula for QCD strong coupling with $N_f=4$ active quark flavours 
and $\Lambda_{\rm QCD}=200$~MeV. To estimate the theoretical uncertainties of our
calculations, we varied the parameter $\xi$ between $1/2$ and $2$.

Our simulations closely follow the experimental setup\cite{JPsiPsi-LHCb-13, JPsiJPsi-LHCb-13}. In particular, the produced $J/\psi$ and $\psi^\prime$ mesons are required 
to be in the region of $2< y <4.5$ 
and $p_T<14$~GeV, where $y$ is the rapidity of a final state particle.
All these conditions have been implemented in our numerical program.

\begin{figure}
\begin{center}
{\includegraphics[width=.32\textwidth]{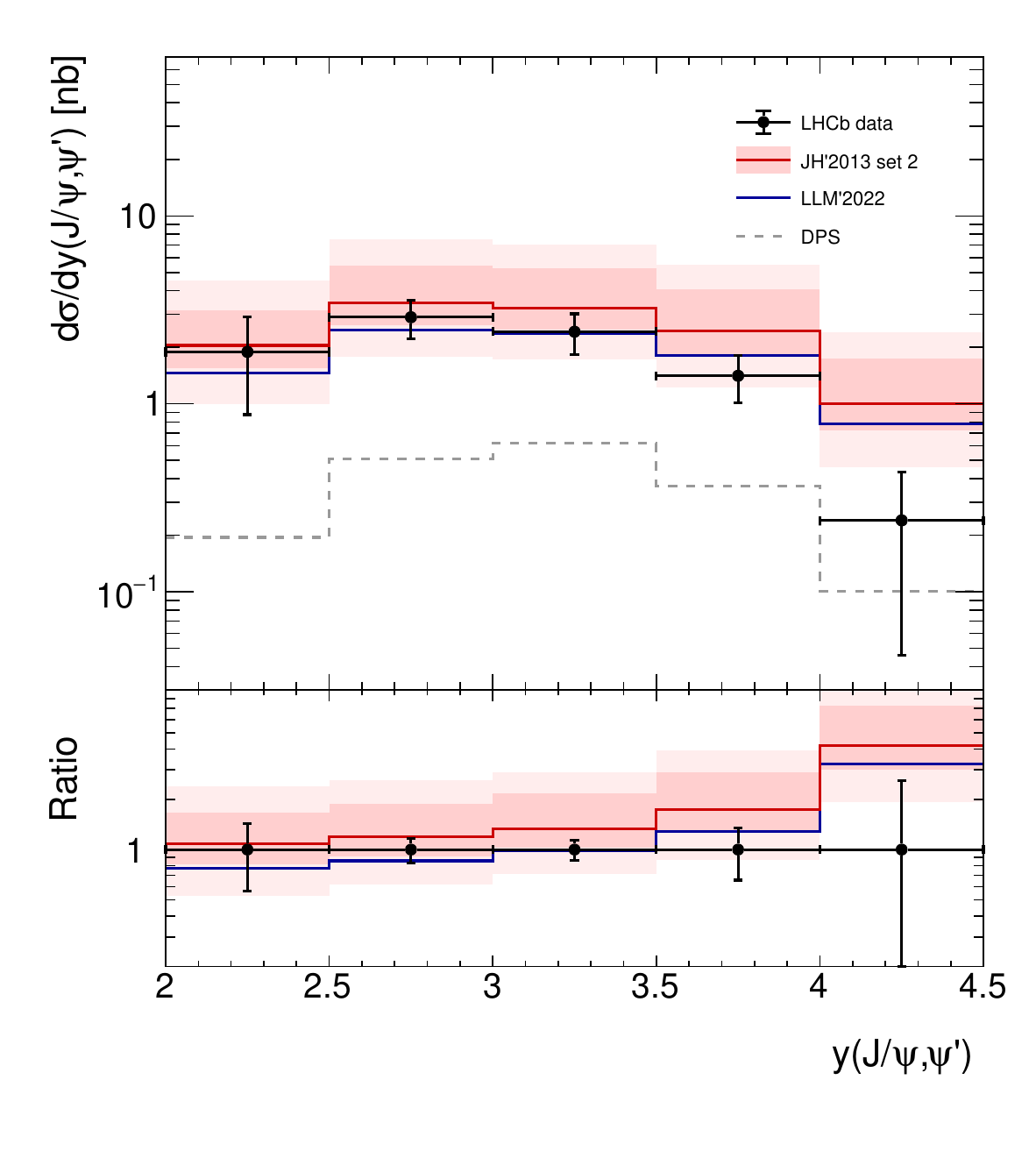}}
{\includegraphics[width=.32\textwidth]{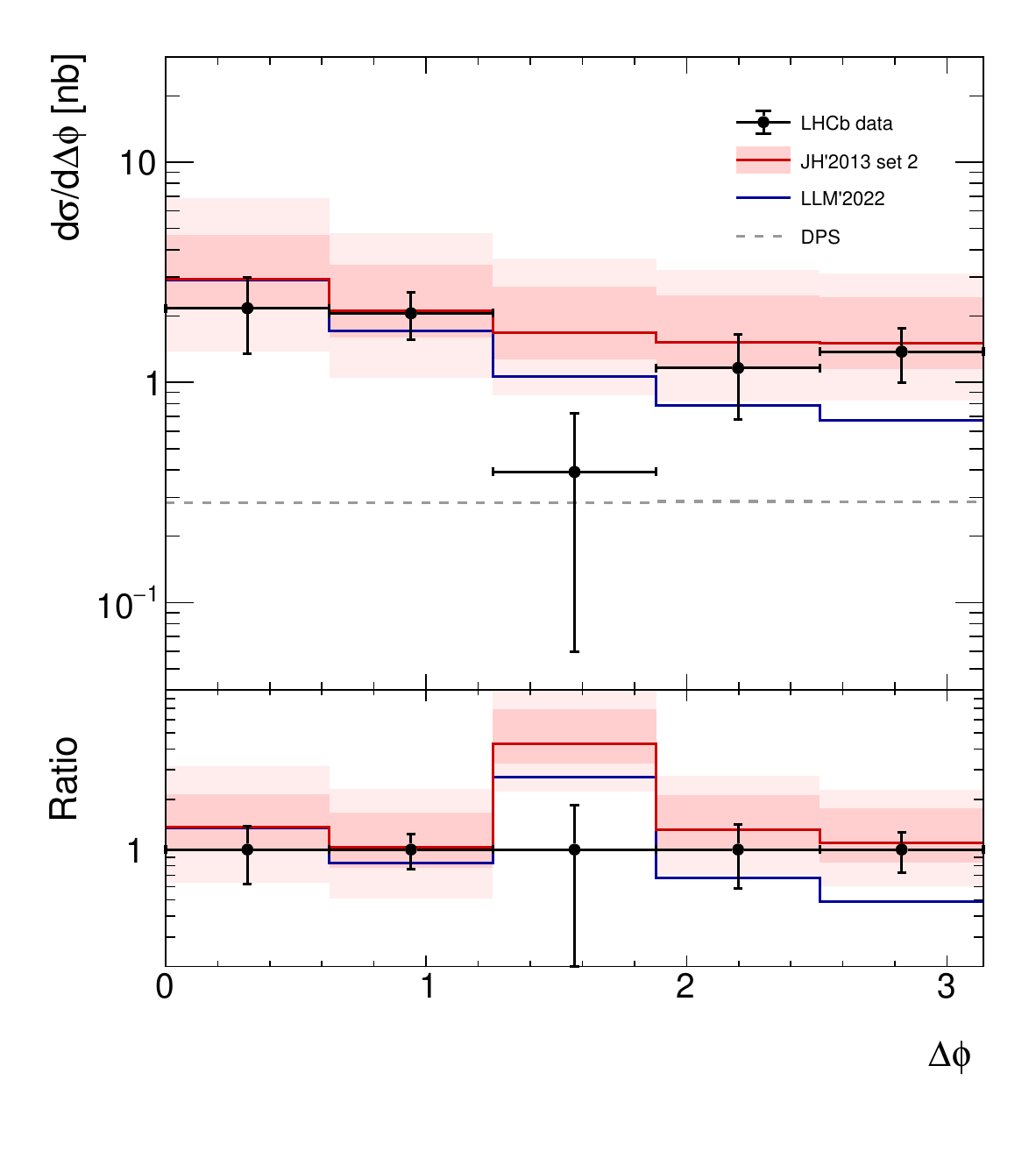}}
{\includegraphics[width=.32\textwidth]{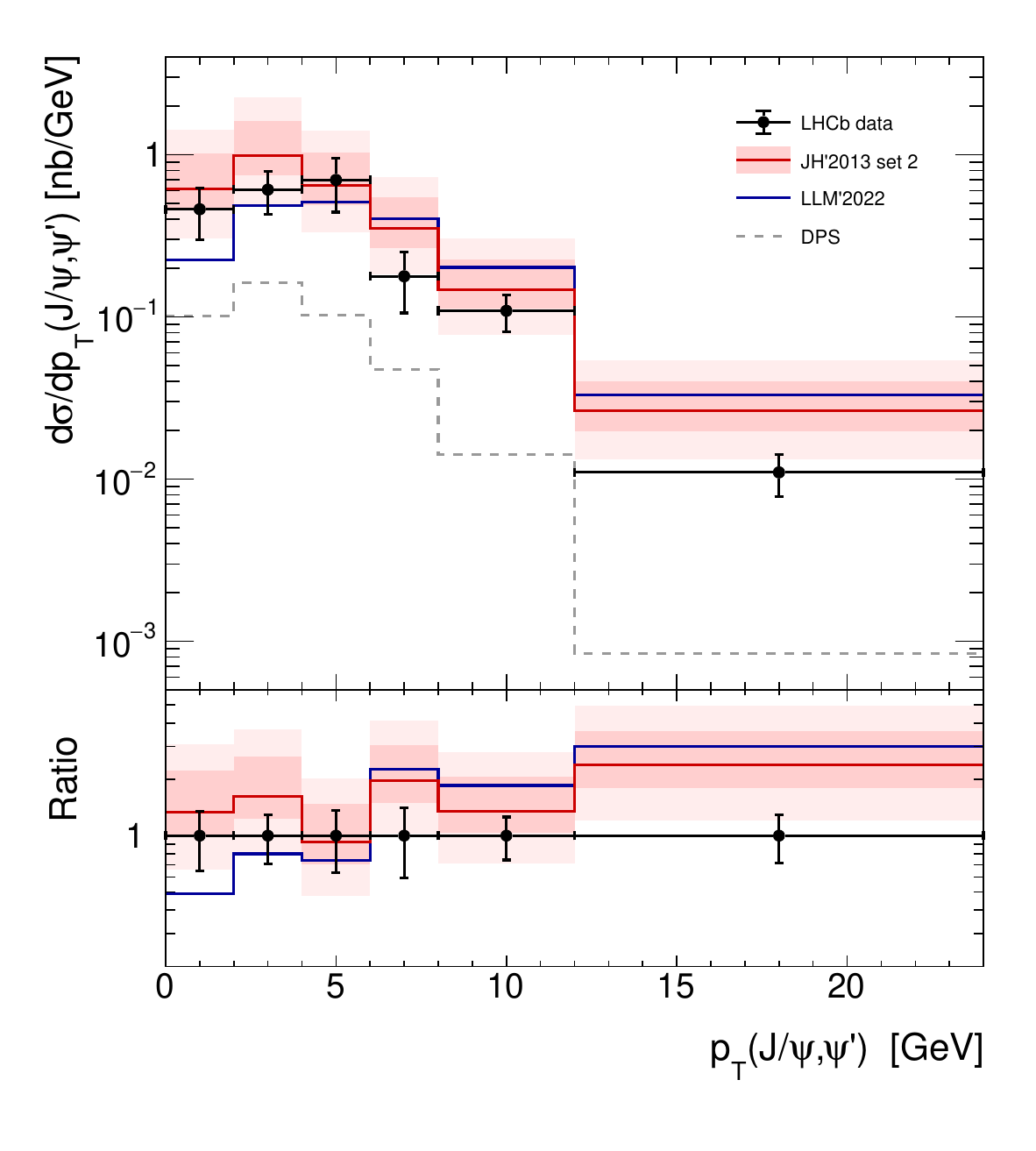}}
{\includegraphics[width=.32\textwidth]{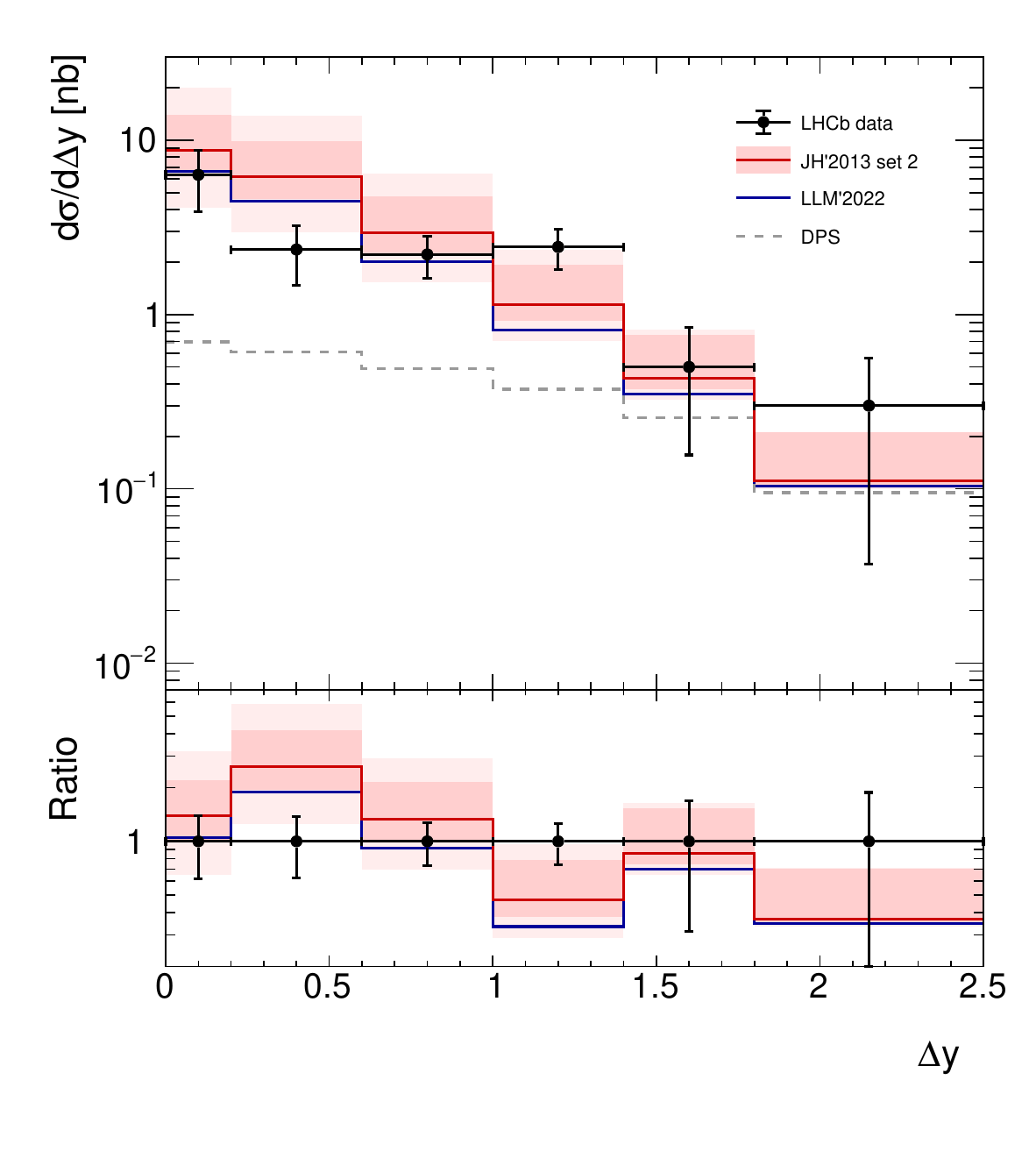}}
{\includegraphics[width=.32\textwidth]{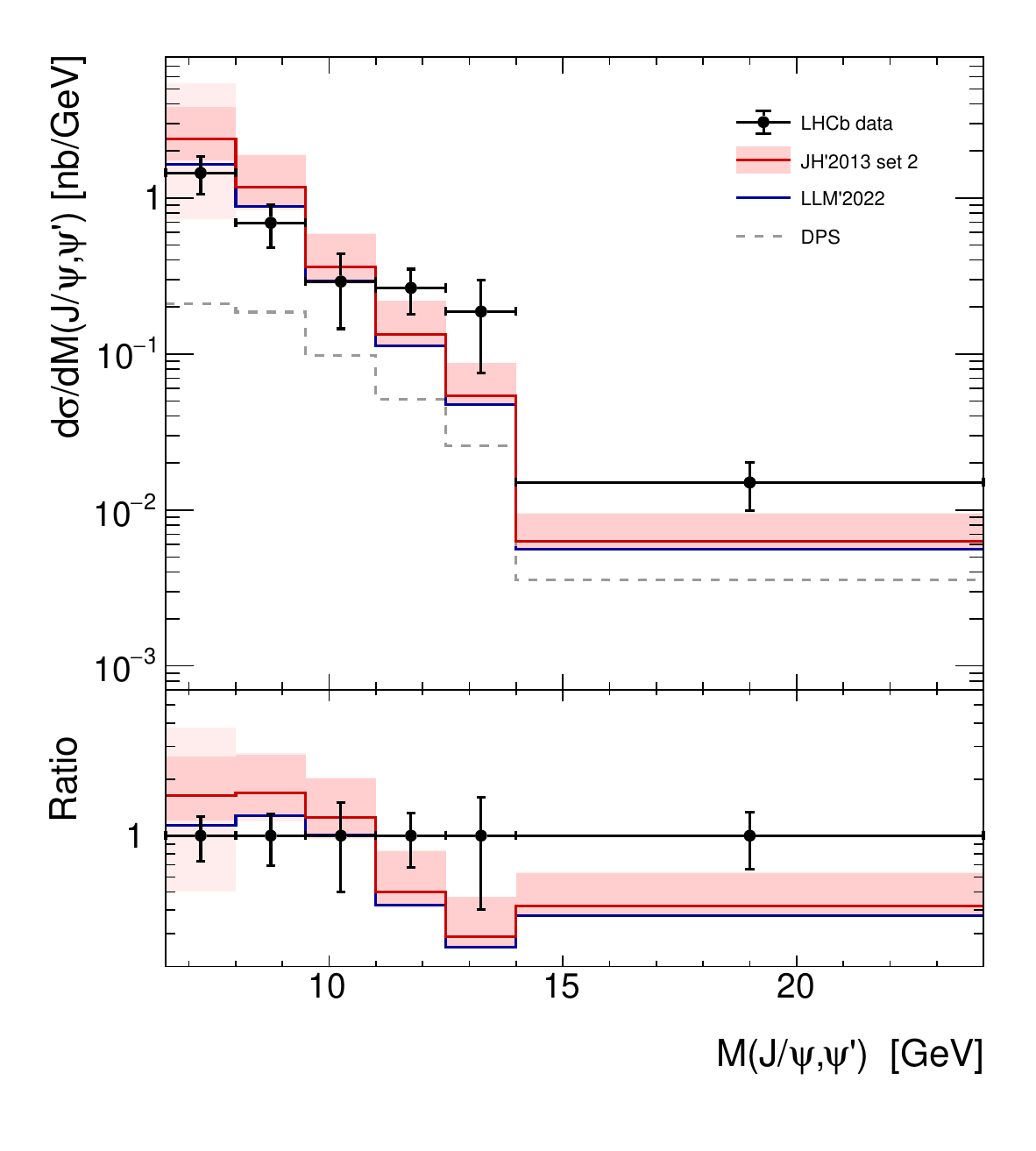}}
\caption{Differential cross sections of associated $J/\psi$ and $\psi^\prime$ production as functions of their rapidity 
difference $\Delta y(J/\psi, \psi^\prime)$, azimuthal angle difference $\Delta\phi(J/\psi, \psi^\prime)$, transverse momentum $p_T(J/\psi, \psi^\prime)$, 
rapidity $y(J/\psi, \psi^\prime)$ and invariant mass $M(J/\psi, \psi^\prime)$. 
Separately shown are the contrbutions from DPS production mechanism.
The data are of LHCb\cite{JPsiPsi-LHCb-13}.}
\label{fig:ATLAS_psi} 
 \end{center}
\end{figure}

The results of our calculations of the
differential cross sections for forward $J/\psi + J/\psi$ and 
$J/\psi + \psi^\prime$ production
are shown in Figs.~\ref{fig:ATLAS_jpsi} and \ref{fig:ATLAS_psi}, respectively.
The theoretical uncertainty bands (represented by the shaded regions)
are shown for JH'2013 set 2 gluon density and
related with scale uncertainties, which have been estimated in a usual way,
by varying the $\xi$ parameter around its default value $\xi = 1$ by a factor of $2$.
This was accompanied with using the JH'2013 set $2+$ and JH'2013 set $2-$ gluon densities in
place of default 
distribution, in accordance with\cite{JH2013}.
We also include the mass uncertainties for CS $J/\psi + \psi^{\prime}$ subprocess 
by varying the c-quark mass between $m(J/\psi)/2 < m_c < m(\psi^{\prime})/2$. 
These effects are added in quadratures with scale uncertainties and represented by the light shaded bands in Fig.~\ref{fig:ATLAS_psi}.
To highlight the role of DPS mechanism, we separately show the corresponding contributions.
One can see that our predictions 
agree well with the latest LHCb data. 
The only exception is seen at larger invariant masses 
$M(J/\psi, J/\psi)$ and $M(J/\psi, \psi^\prime)$,
where the calculated cross sections systematically underestimate the experimental results.
Note that still additional contributions related to multiple gluon radiation 
in the initial gluon evolution cascade\footnote{Such terms 
are extremely important at central rapidities\cite{DoubleJPsi-we2, DoubleJPsi-we3}.} could play a role at large $M(J/\psi, J/\psi)$ and $M(J/\psi, \psi^\prime)$.
However, an accurate account of all these terms needs rather lengthy numerical calculations and is therefore left out of our present scope.

We find special interest in the correlation observables
$\Delta \phi(J/\psi,\psi^\prime)$, $\Delta \phi(J/\psi,J/\psi)$
and in the $p_T$ balance observables 
$p_T(J/\psi,J/\psi)$ and $p_T(J/\psi,\psi^\prime)$
as they are known to be particularly sensitive to the non-collinear gluon 
evolution dynamics in a proton (see, for example,\cite{AzimuthalCorrelations} and references therein).
We point out a good description (within the uncertainties) of all these distributions
with the JH'2013 set 2 gluon density.
The LLM-based predictions, in general, tend to slightly underestimate the LHCb data\cite{JPsiPsi-LHCb-13, JPsiJPsi-LHCb-13} and JH'2013 set~2 predictions. 
In fact, the measured fiducial cross sections of forward
$J/\psi + J/\psi$ and $J/\psi + \psi^\prime$ production
are $\sigma(J/\psi + J/\psi) = 16.36 \pm 0.92$~nb and $\sigma(J/\psi + \psi^\prime) = 4.5 \pm 0.8$~nb.
They can be compared with the JH'2013 set 2 and LLM'2022 results, 
which are $\sigma^{\rm JH}(J/\psi + J/\psi) = 16.2^{+7.2}_{-3.5}$ (scale unc.)~nb, 
$\sigma^{\rm JH}(J/\psi + \psi^\prime) = 6.1^{+3.7}_{-1.5}$ (scale unc.)$^{+6.2}_{-2.6}$ 
(mass unc.)~nb
and $\sigma^{\rm LLM}(J/\psi + J/\psi) = 12.1^{+4.9}_{-2.3}$ (scale unc.)~nb, 
$\sigma^{\rm LLM}(J/\psi + \psi^\prime) = 4.5^{+2.5}_{-0.9}$ (scale unc.)$^{+4.3}_{-1.8}$ (mass unc.)~nb, respectively.
Moreover, one can see a clear difference between the JH'2013 set 2 and LLM'2022 predictions
in the shape of azimuthal angle correlations and transverse momenta correlations of the
produced mesons.
It demonstrates again that such observables could be useful to 
discriminate the different TMD evolution scenarios.

\begin{figure}
\begin{center}
{\includegraphics[width=.32\textwidth]{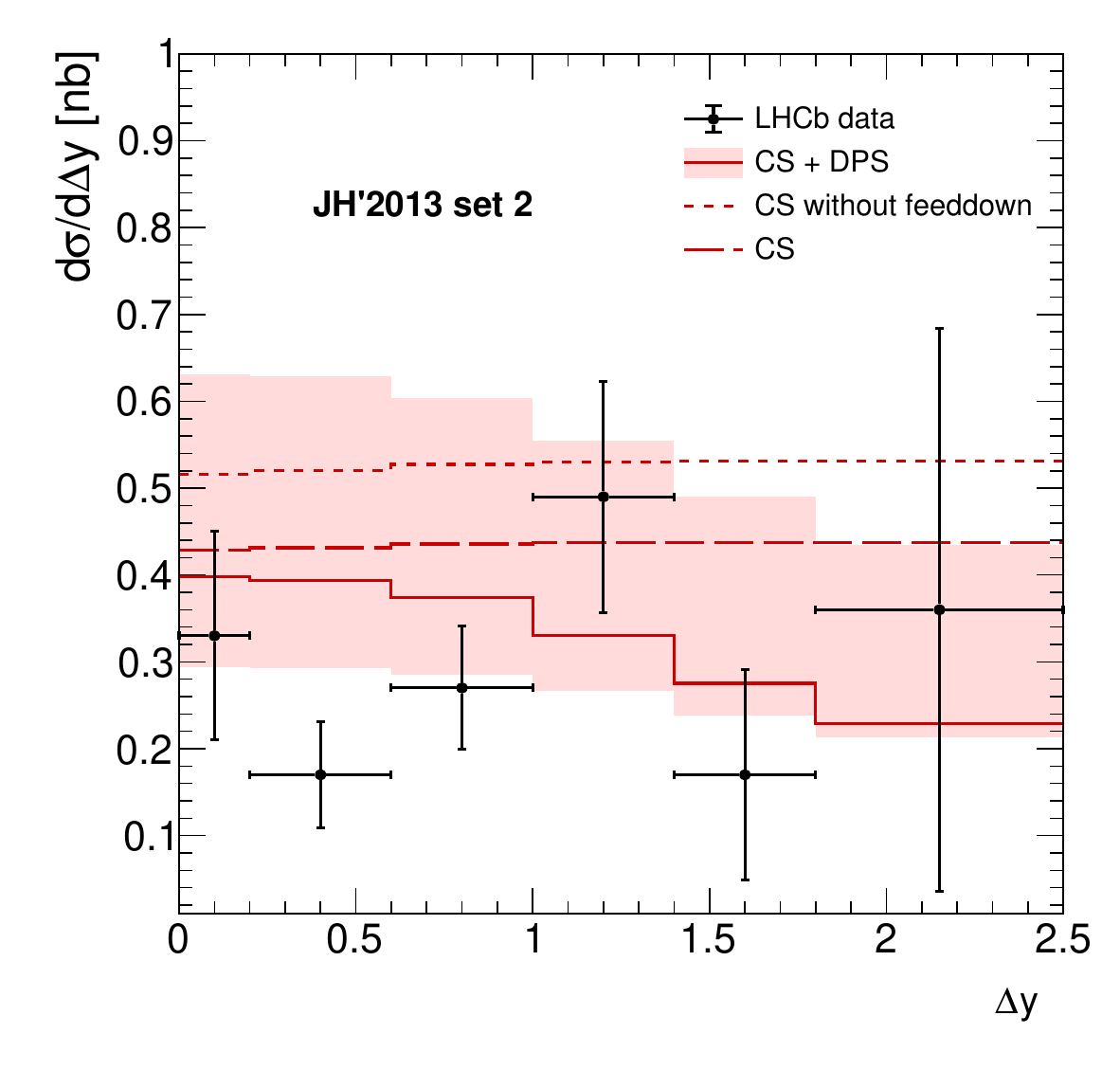}}
{\includegraphics[width=.32\textwidth]{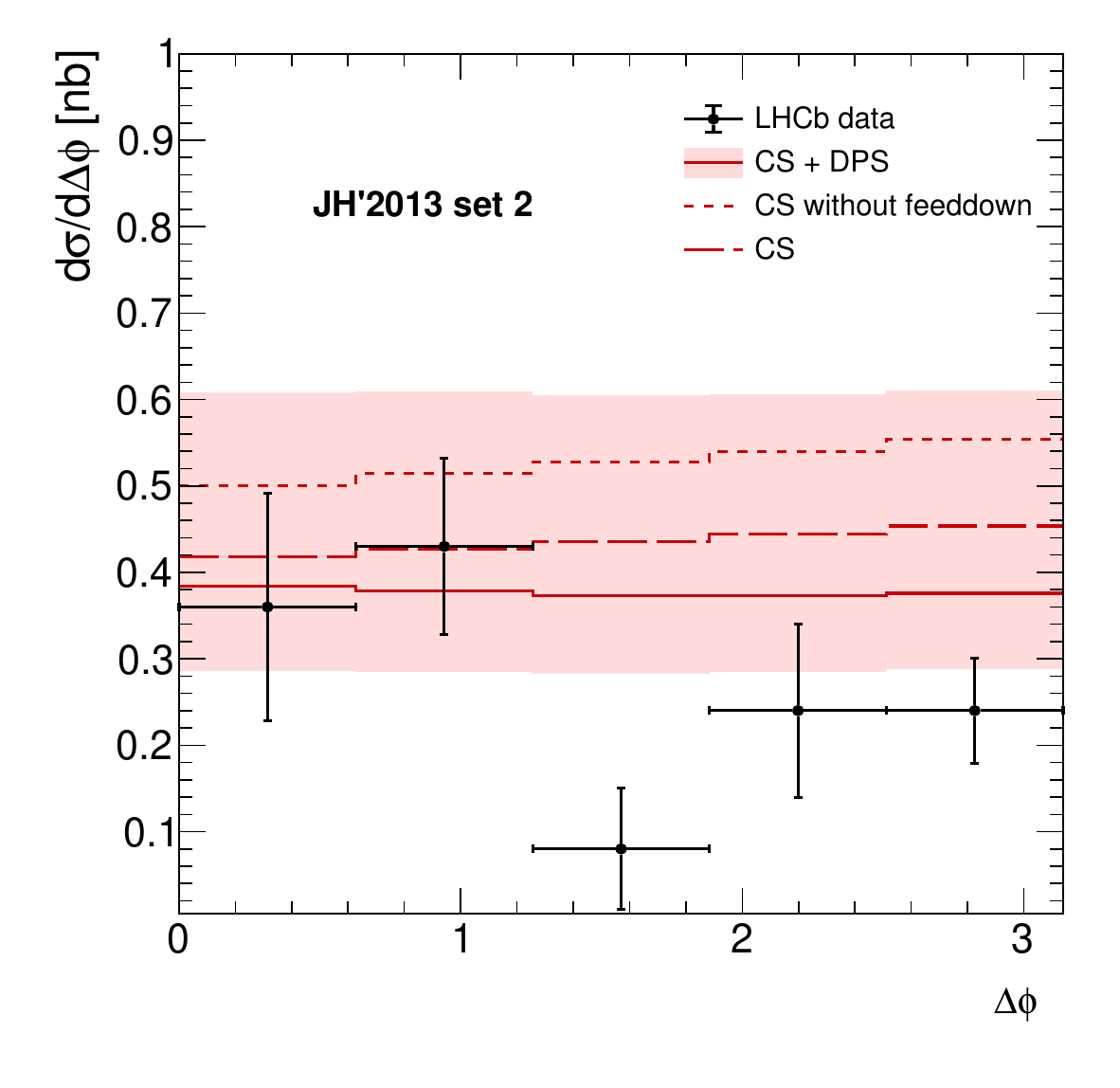}}
{\includegraphics[width=.32\textwidth]{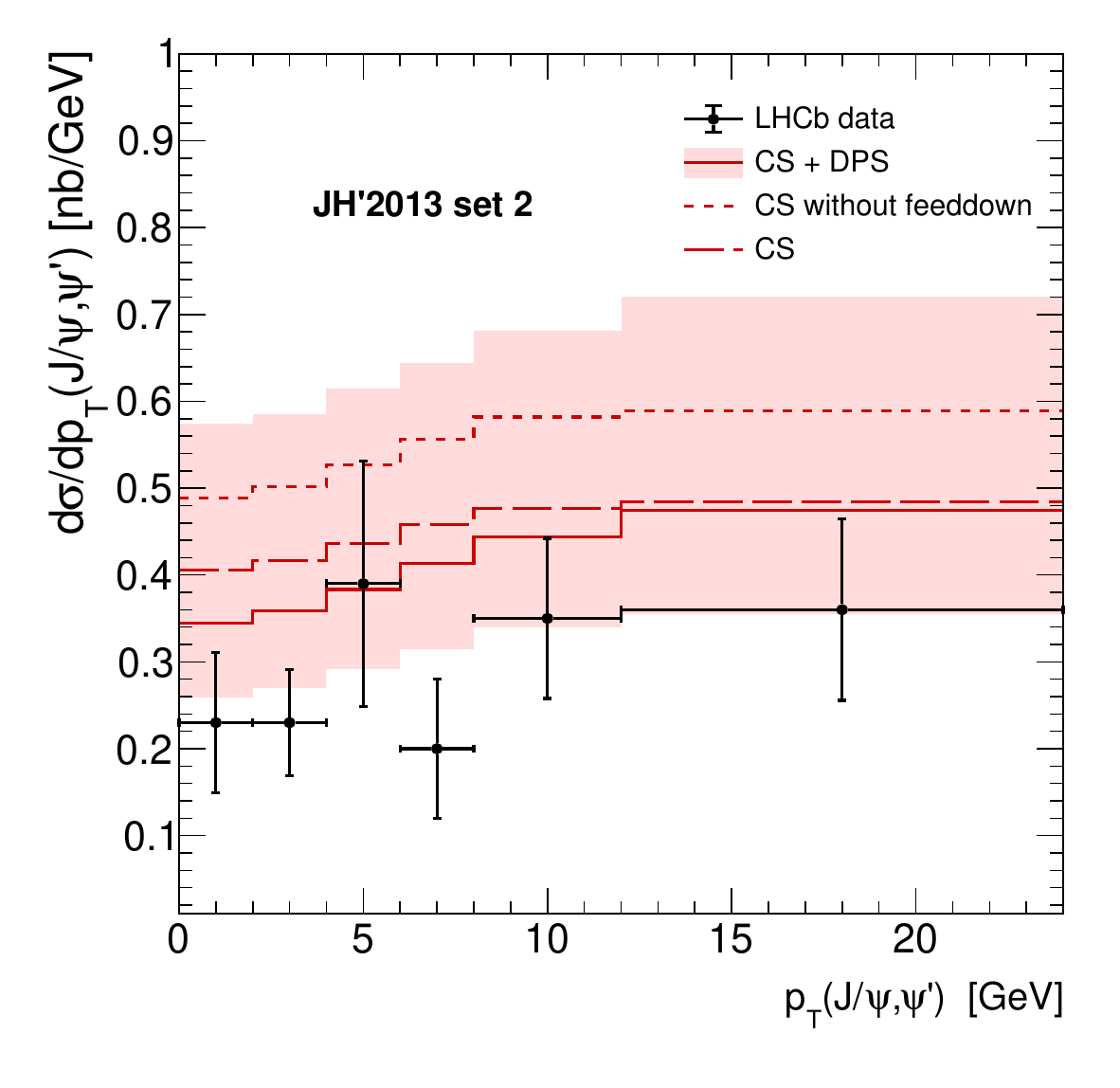}}
{\includegraphics[width=.32\textwidth]{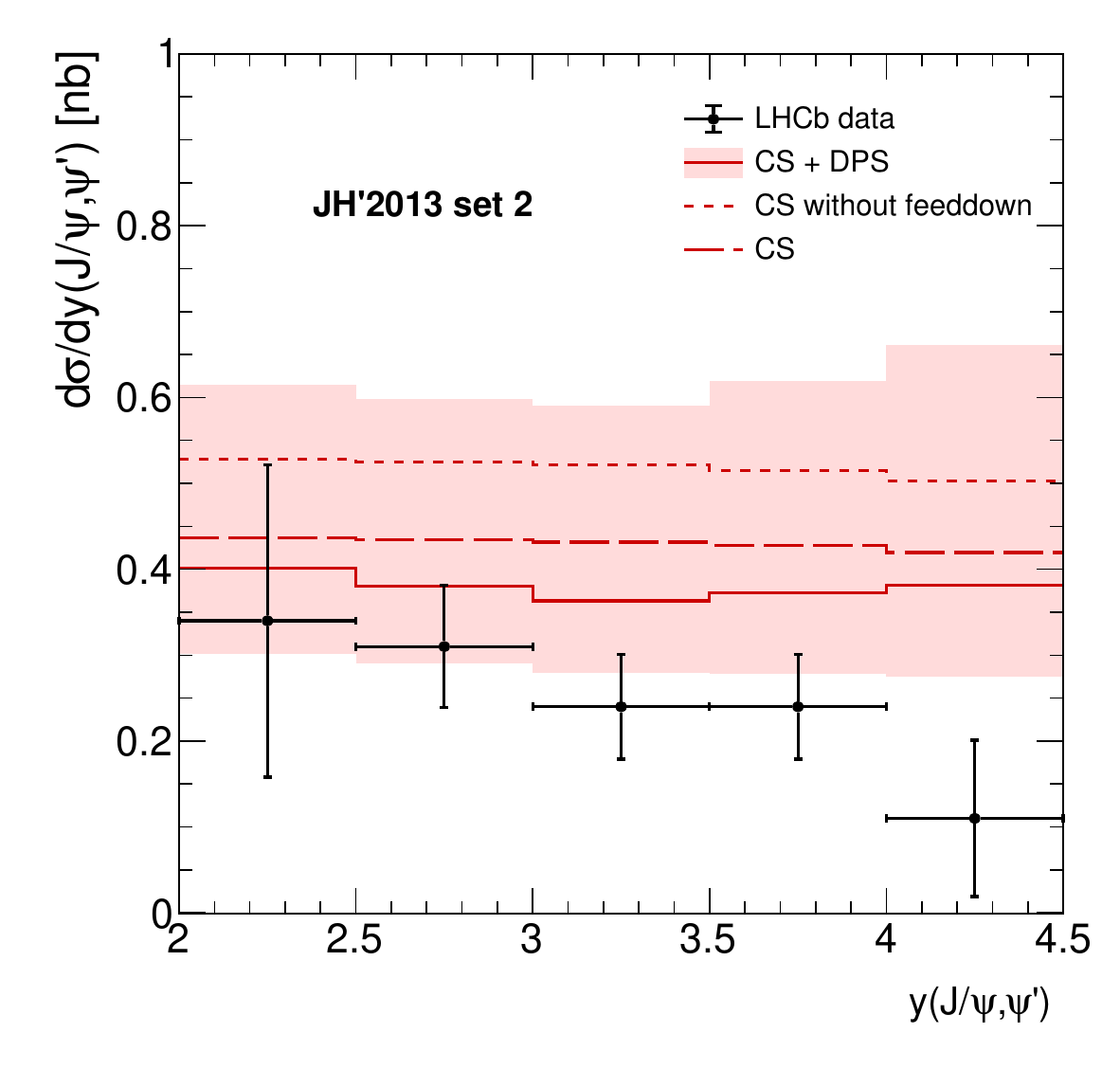}}
\caption{Relative production rate $\sigma(J/\psi + \psi^\prime)/\sigma(J/\psi + J/\psi)$
calculated at $\sqrt s = 13$~TeV as a function of several kinematic variables.
The data are of LHCb\cite{JPsiPsi-LHCb-13}.}
\label{fig:LHCb_ratio} 
 \end{center}
\end{figure}

The estimated contributions from the DPS production mechanism are typically small and
only become important in some specific regions like low $p_T(J/\psi, J/\psi)$, $p_T(J/\psi, \psi^\prime)$ 
and large $M(J/\psi, J/\psi) \geq 15$~GeV, $M(J/\psi,\psi^{\prime}) \geq 15$~GeV, $|\Delta y(J/\psi,J/\psi)| \geq 1.5$, $|\Delta y(J/\psi,\psi^\prime)| \geq 1.5$.
Our calculations clearly show that taking these terms into account improves the quality of the data description and becomes necessary in the forward kinematic region.
The role of DPS and feeddown contributions is exhibited in Fig.~\ref{fig:LHCb_ratio},
where the ratio $\sigma(J/\psi + \psi^\prime)/\sigma(J/\psi + J/\psi)$
is shown as a function of the different kinematic variables.
We find that the relative production rate is very
sensitive to the DPS and feeddown yields\footnote{Note that only the SPS terms are affected
by the choice of TMD gluon density in a proton. The DPS predictions are stable, because when we switch to
a different set of TMD gluons we have to accordingly change the LDME values, thus making the net result the same.}.
Finally, we can conclude that the first LHCb measurement\cite{JPsiPsi-LHCb-13} of the associated $J/\psi + \psi^\prime$
production and the very recent experimental data\cite{JPsiJPsi-LHCb-13} on $J/\psi + J/\psi$
production at $\sqrt s = 13$~TeV can be well described by the color singlet SPS mechanism 
and DPS contributions calculated with the standard choice of $\sigma_{\rm eff}$ parameter.

{\it Acknowledgements.} A.V.L. and M.A.M. would like to thank School of Physics and Astronomy, Sun Yat-sen University (Zhuhai, China)
for warm hospitality. P.M.Z. was partially supported by the National Natural Science Foundation of China (Grant No. $12375084$).




\bibliography{psipsi7-letter-sp}

\begin{thebibliography}{10}

\bibitem{JPsiPsi-LHCb-13}
LHCb~Collaboration{,} arXiv:2311.15921 [hep{-}ex].

\bibitem{JPsiJPsi-LHCb-13}
LHCb Collaboration{,} JHEP {\bf 03}{,}~088 (2024).

\bibitem{NRQCD-1}
G.~Bodwin{,} E.~Braaten{,} G.~Lepage{,} Phys. Rev. D {\bf 51}{,}~1125 (1995).

\bibitem{NRQCD-2}
P.~Cho{,} A.K.~Leibovich{,} Phys. Rev. D {\bf 53}{,} 150 (1996);\\ P.~Cho{,}
  A.K.~Leibovich{,} Phys. Rev. D {\bf 53}{,}~6203 (1996).

\bibitem{DoubleJPsi-Lansberg2}
J.{-}P.~Lansberg{,} H.{-}S.~Shao{,} N.~Yamanaka{,} Y.{-}J.~Zhang{,} Eur. Phys.
  J. C {\bf 79}{,}~1006 (2019).

\bibitem{DoubleJPsi-Gen1}
H.{-}S.~Shao{,} Comput. Phys. Commun. {\bf 184}{,}~2562 (2013).

\bibitem{DoubleJPsi-Gen2}
H.{-}S.~Shao{,} Comput. Phys. Commun. {\bf 198}{,}~238 (2016).

\bibitem{DoubleJPsi-Kniehl1}
Z.{-}G. He{,} B.A.~Kniehl{,} Phys. Rev. Lett. {\bf 115}{,}~022002 (2015).

\bibitem{DoubleJPsi-Lansberg1}
J.{-}P.~Lansberg{,} H.{-}S.~Shao{,} Phys. Rev. Lett. {\bf 111}{,}~122001
  (2013).

\bibitem{kt-factorization}
L.V.~Gribov{,} E.M.~Levin{,} M.G.~Ryskin{,} Phys. Rep. {\bf 100}{,} 1 (1983){;}
  \\ E.M.~Levin{,} M.G.~Ryskin{,} Yu.M.~Shabelsky{,} A.G.~Shuvaev{,} Sov. J.
  Nucl. Phys. {\bf 53}{,}~657 (1991).

\bibitem{HighEnergyFactorization}
S.~Catani{,} M.~Ciafaloni{,} F.~Hautmann{,} Nucl. Phys. B {\bf 242}{,} 97
  (1990){;} \\ S.~Catani{,} M.~Ciafaloni{,} F.~Hautmann{,} Nucl. Phys. B {\bf
  366}{,} 135 (1991){;} \\ J.C.~Collins{,} R.K.~Ellis{,} Nucl. Phys. B {\bf
  360}{,}~3 (1991).

\bibitem{TMD-Review}
R.~Angeles-Martinez{,} A. Bacchetta{,} I.I. Balitsky{,} D. Boer{,} M.
  Boglione{,} R. Boussarie{,} F.A. Ceccopieri{,} I.O. Cherednikov{,} P.
  Connor{,} M.G. Echevarria{,} G. Ferrera{,} J. Grados Luyando{,} F.
  Hautmann{,} H. Jung{,} T. Kasemets{,} K. Kutak{,} J.P. Lansberg{,} A.
  Lelek{,} G.I. Lykasov{,} J.D. Madrigal Martinez{,} P.J. Mulders{,} E.R.
  Nocera{,} E. Petreska{,} C. Pisano{,} R. Placakyte{,} V. Radescu{,} M.
  Radici{,} G. Schnell{,} I. Scimemi{,} A. Signori{,} L. Szymanowski{,} S.
  Taheri Monfared{,} F.F.~Van der Veken{,} H.J.~van Haevermaet{,} P. Van
  Mechelen{,} A.A. Vladimirov{,} S. Wallon{,} Acta Phys. Polon. B {\bf 46}{,}
  2501~(2015).

\bibitem{DoubleJPsi-we1}
S.P.~Baranov{,} Phys. Rev. D {\bf 84}{,}~054012 (2011).

\bibitem{DoubleJPsi-we2}
A.A.~Prokhorov{,} A.V.~Lipatov{,} M.A.~Malyshev{,} S.P.~Baranov{,} Eur. Phys.
  J. C {\bf 80}{,}~1046 (2020).

\bibitem{DoubleJPsi-we3}
S.P.~Baranov{,} A.V.~Lipatov{,} A.A.~Prokhorov{,} Phys. Rev. D {\bf
  106}{,}~034020 (2022).

\bibitem{DoubleJPsi-Kniehl2}
Z.{-}G. He{,} B.A.~Kniehl{,} M.A.~Nefedov{,} V.A.~Saleev{,} Phys. Rev. Lett.
  {\bf 123}{,}~162002 (2019).

\bibitem{BFKL}
E.A.~Kuraev{,} L.N.~Lipatov{,} V.S.~Fadin{,} Sov. Phys. JETP {\bf 44}{,} 443
  (1976); \\ E.A.~Kuraev{,} L.N.~Lipatov{,} V.S.~Fadin{,} Sov. Phys. JETP {\bf
  45}{,} 199 (1977); \\ I.I.~Balitsky{,} L.N.~Lipatov{,} Sov. J. Nucl. Phys.
  {\bf 28}{,}~822 (1978).

\bibitem{CCFM}
M.~Ciafaloni{,} Nucl. Phys. B {\bf 296}{,} 49 (1988); \\ S.~Catani{,}
  F.~Fiorani{,} G.~Marchesini{,} Phys. Lett. B {\bf 234}{,} 339 (1990); \\
  S.~Catani{,} F.~Fiorani{,} G.~Marchesini{,} Nucl. Phys. B {\bf 336}{,} 18
  (1990); \\ G.~Marchesini{,} Nucl. Phys. B {\bf 445}{,}~49 (1995).

\bibitem{ProjectionOperators-1}
C.-H. Chang{,} Nucl. Phys. B {\bf 172}{,}~425 (1980).

\bibitem{ProjectionOperators-2}
E.L. Berger{,} D. Jones{,} Phys. Rev. D {\bf 23}{,}~1521 (1981).

\bibitem{ProjectionOperators-3}
R.~Baier{,} R. R\"uckl{,} Phys. Lett. B {\bf 102}{,}~364 (1981).

\bibitem{ProjectionOperators-4}
H.~Krasemann{,} Z. Phys. C {\bf 1}{,}~189 (1979).

\bibitem{ProjectionOperators-5}
G.~Guberina{,} J. K\"uhn{,} R. Peccei{,} R. R\"uckl{,} Nucl. Phys. B {\bf
  174}{,}~317 (1980).

\bibitem{JH2013}
F.~Hautmann{,} H. Jung{,} Nucl. Phys. B {\bf 883}{,}~1 (2014).

\bibitem{LLM-2022}
A.V.~Lipatov{,} G.I.~Lykasov{,} M.A.~Malyshev{,} Phys. Rev. D {\bf
  107}{,}~014022 (2023).

\bibitem{Motyka-photon}
K.~Golec-Biernat{,} L. Motyka{,} T. Stebel{,} Phys. Rev. D {\bf 103}{,}~034013
  (2021).

\bibitem{LMJ-PP}
A.V.~Lipatov{,} M.A.~Malyshev{,} H.~Jung{,} Phys. Rev. D {\bf 100}{,}~034028
  (2019).

\bibitem{LM-Higgs}
A.V.~Lipatov{,} M.A.~Malyshev{,} Phys. Rev. D {\bf 103}{,}~094021 (2020).

\bibitem{LLM-FL}
A.V.~Lipatov{,} G.I.~Lykasov{,} M.A.~Malyshev{,} Phys. Lett. B {\bf
  839}{,}~137780 (2023).

\bibitem{LLM-photon}
A.V.~Lipatov{,} M.A.~Malyshev{,} Phys. Rev. D {\bf 108}{,}~014022 (2023).

\bibitem{ModifiedSoftQuarkGluonStringModel-1}
V.A. Bednyakov{,} G.I. Lykasov{,} V.V. Lyubushkin{,} Europhys. Lett. {\bf
  92}{,}~31001 (2010).

\bibitem{ModifiedSoftQuarkGluonStringModel-2}
V.A. Bednyakov{,} A.A. Grinyuk{,} G.I. Lykasov{,} M. Poghosyan{,} Int. J. Mod.
  Phys. A {\bf 27}{,}~1250042 (2012).

\bibitem{TMDLib2}
N.A. Abdulov{,} A. Bacchetta{,} S.P. Baranov{,} A. Bermudez Martinez{,} V.
  Bertone{,} C. Bissolotti{,} V. Candelise{,} L.I. Estevez Banos{,} M. Bury{,}
  P.L.S. Connor{,} L. Favart{,} F. Guzman{,} F. Hautmann{,} M. Hentschinski{,}
  H. Jung{,} L. Keersmaekers{,} A.V. Kotikov{,} A. Kusina{,} K. Kutak{,} A.
  Lelek{,} J. Lidrych{,} A.V. Lipatov{,} G.I. Lykasov{,} M.A. Malyshev{,} M.
  Mendizabal{,} S. Prestel{,} S. Sadeghi Barzani{,} S. Sapeta{,} M. Schmitz{,}
  A. Signori{,} G. Sorrentino{,} S. Taheri Monfared{,}~A. van Hameren{,}
  A.M.~van Kampen{,} M. Vanden Bemden{,} A. Vladimirov{,} Q. Wang{,} H. Yang{,}
  Eur. Phys. J. C {\bf 81}{,} 752~(2021).

\bibitem{PEGASUS}
A.V.~Lipatov{,} M.A.~Malyshev{,} S.P.~Baranov{,} Eur. Phys. J. C {\bf
  80}{,}~330 (2020).

\bibitem{DPS-forward1}
S.P.~Baranov{,} A.M.~Snigirev{,} N.P.~Zotov{,} Phys. Lett. B {\bf 705}{,}~116
  (2011).

\bibitem{DPS-forward2}
C.H.~Kom{,} A.~Kulesza{,} W.J.~Stirling{,} Phys. Rev. Lett. B {\bf
  107}{,}~082002 (2011).

\bibitem{DPS-forward3}
S.P.~Baranov{,} A.M.~Snigirev{,} N.P.~Zotov{,} A.~Szczurek{,} W.~Sch\"afer{,}
  Phys. Rev. D {\bf 87}{,}~034035 (2013).

\bibitem{TMD-charmonia}
S.P.~Baranov{,} A.V. Lipatov{,} Phys. Rev. D {\bf 100}{,}~114021 (2019).

\bibitem{DPS-DQ1}
D0~Collaboration{,} Phys. Rev. D {\bf 90}{,}~111101(R) (2014).

\bibitem{DPS-DQ2}
D0~Collaboration{,} Phys. Rev. Lett {\bf 116}{,}~082002 (2016).

\bibitem{DPS-DQ3}
ATLAS Collaboration{,} Eur. Phys. J. C {\bf 77}{,}~76 (2017).

\bibitem{DPS-DQ4}
CMS Collaboration{,} JHEP {\bf 1705}{,}~013 (2017).

\bibitem{PDG}
PDG Collaboration{,} Prog. Theor. Exp. Phys. 2022{,}~083C01 (2022).

\bibitem{AzimuthalCorrelations}
S.P.~Baranov{,} A.V.~Lipatov{,} M.A.~Malyshev{,} Eur. Phys. J. C {\bf
  78}{,}~820 (2018).

\end{thebibliography}

\end{document}